# Voronoi Entropy and long-range order of 2D point sets

**Edward Bormashenko[1*], Shraga Shoval[2], Mark Frenkel,[1] Michael Nosonovsky[3*]**

[1] Department of Chemical Engineering, Ariel University, Ariel, POB 3, 407000, Israel
[2] Department of Industrial Engineering and Management, Faculty of Engineering, Ariel University, P.O.B. 3, Ariel 407000, Israel
[3] Department of Mechanical Engineering, University of Wisconsin-Milwaukee, Milwaukee, WI 53211, USA
*Correspondence: edward@ariel.ac.il

**Abstract**

The Voronoi Entropy (VE) and the continuous measure of symmetry (CSM) characterize the orderliness of a set of points on a 2D plane. The Voronoi entropy is the Shannon entropy of the Voronoi tessellation of the plane into polygons, quantifying the diversity of polygons. The VE is widely used to study the self-assembly of colloidal, supramolecular, and other systems. The value of VE changes from $S = 0$ for a completely ordered system, built of polygons with an equal number of sides, to $S = 1.690 \pm 0.001$ for a random set of points. While the VE takes into account only neighboring polygons, covering the 2D plane imposes constraints on the number of polygons and the number of edges in polygons. Consequently, unlike the conventional Shannon Entropy, the VE captures some long-range order properties of the system. We calculate the VE for several hyperuniform sets of points and compare it with the values of exponents of collective density variables characterizing long-range correlations in the system. We show that the VE correlates with the latter up to a certain saturation level, after which the value of the VE falls to $S = 0$, and we explain this phenomenon.

### Introduction

Characterizing spatial patterns of particles in many-body systems is important for various applications such as colloidal crystals [1], droplet clusters [2-3], molecules [4], and supramolecular systems [5]. One of the challenges is understanding the complex relationships between the long-range (global) and short-range (local) order in such systems. These relationships are often elusive when quantitative measures of orderliness are used. Among various quantitative characteristics of the orderliness of such systems, the Voronoi Entropy (VE) [2], Continuous symmetry measure (CSM) [6], and collective density variables [7] have been suggested.

The Voronoi tessellation of a set of points and calculation of the VE is a common method to estimate quantitatively the orderliness of various 2D systems, such as self-assembled droplet clusters, colloidal crystals, supramolecular systems, polymer breath figures. An infinite plane with a discrete set of seed (or "nuclei") points is partitioned into polygons based on the distance to each point to form a Voronoi tessellation. For every seed, there is a corresponding polygonal region consisting of all points closer to that seed than to any other. Consequently, the number of edges of every polygon obtained by the Voronoi tessellation depends on the neighboring points, in other words, it is local. This makes VE a measure of orderliness that takes into account the short-range order in a set of points, i.e., only neighboring points are accounted for. On the other hand, the Voronoi tessellation reflects the general topological properties of the surface on which it is constructed. The Voronoi tessellation, as any other planar cellular pattern, obeys the seminal Euler topological equation, to be discussed in detail in the Methods Section. Thus, the Voronoi tessellation on one hand is defined by the local properties of the pattern, and on the other hand, it follows the general topological properties of the surface, on which it is built.

Now consider the quantitative measures of "orderliness" of the Voronoi tessellations. It should be emphasized that the notion of "orderliness" has a fine structure; in other words, it cannot be quantified with a single numerical parameter [8]. We studied two parameters quantifying "orderliness" of hyperuniform sets of points, namely: the Voronoi/Shannon entropy and continuous measure of symmetry (abbreviated CSM) of the addressed 2D patterns. It was demonstrated recently that the maxima and minima of the VE and CSM of the 2D patterns are not always well correlated; moreover, in certain cases, the maxima of the CSM may correspond to the minima of the VE [8]. Let us start with the Voronoi/Shannon entropy of the patterns.

A Voronoi tessellation is built of polygons with the number of edges from $n=3$ to $n\rightarrow\infty$. The VE is defined as the Shannon entropy of the Voronoi tessellation calculated as

$$S = -\sum_{n=3}^{\infty} p_n \ln(p_n) \qquad (1)$$

where $p_n$ is the fraction of polygons with $n$ sides or edges (also called the coordination number of the polygon) in a given Voronoi diagram. The VE, supplied by Eq. 1 quantifies the diversity of polygons in a given tessellation, which can be *cum grano*

*salis* seen as a measure of orderliness of the pattern. An important property of the VE is that it does not depend on the number of points in the system, since it employs only normalized quantities, $p_n$, thus, it may be seen as an intensive property of a given pattern. For a perfectly "ordered" tessellation, for example, one built only by hexagons, the value is $S = 0$. On the other hand, for a random tessellation (**Table 1**), the value is $S \approx 1.69$.

The Continuous Symmetry Measure (CSM) [6] for a set of *m* points, denoted by $\psi(G)$ is based on measuring the deviation of an approximately symmetrical set from the exact symmetry, usually employing the sum of squares of the distances between the symmetric and disturbed position of the points

$$\psi = \frac{1}{mR_g^2}\sum_{i=1}^{m}\left|\vec{M_i} - \vec{M_{0i}}\right|^2 \qquad (2)$$

where $\vec{M_i}$ is the position of the *i*-th point, and $\vec{M_{0i}}$ is its position in the symmetric set subject to some symmetry group *G*, and $R_g$ is the radius of gyration about the center of mass of the symmetric set. The CSM is often interpreted as a "minimal effort" required for the transformation of an original shape into a symmetric one. For a Voronoi tessellation, CSM can be defined as the average of all polygonal cells [8].

The CSM was successfully used for the study of approximately symmetric molecules. A comparison of the VE and the CSM was conducted by Frenkel et al. (2021) for levitating self-assembled droplet clusters [8]. They showed that the maxima and minima of the VE and CSM are not always well correlated. Symmetry and orderliness of 2D patterns could not be quantified with a single mathematical measure.

Collective density variables are used to characterize hyperuniform systems with long-range order, which, however, correlate weaker than perfectly ordered crystals [7]. Hyperuniformity is a relatively new concept in statistical physics based on the long-range order in the material [9]. For disordered many-particle systems, such as an ideal gas, the variance of density (or of the number of points representing molecules per unit volume or area) is proportional to the volume (in the 3D case) or to the area (in the 2D case) of the system. An equivalent definition implies that the structure factor vanishes at the long-wavelength limit, $\lim s(k) = 0$ or $s(k) \sim |k|^\gamma$ for $k \to 0$, where $k$ is the wavenumber and $\gamma$ is an exponent. That means that there is no correlation between the occurrence of the points separated by a long distance, $1/k$. For $0 < \gamma < 1$, the long-range correlation vanishes faster than for a crystalline pattern, but slower than for a completely random set of points (i.e., $\gamma = 1$) [7, 9].

Since the exponents of hyperuniform systems are based on the long-range order, while VE is based on the short-range order, the correlation between the two is not trivial. In this paper, we will investigate such a correlation.

**Method**

*Voronoi/Shannon entropy*

The first step of calculating the VE of a set of points on a 2D surface is a generation of the Voronoi Tessellation. A Voronoi tessellation is built of polygons with the number of edges from $n=3$ to $n\rightarrow\infty$, although usually, the number of edges is small; thus, the maximum observed number of edges in a random system that has been reported in the literature is $n=15$ [10, 11]. The relative fraction of polygons with $n$ edges is given by $0 \leq p_n < 1$. From the Euler topological equation, $V - N + F = 2$, which relates the total number of vertices, $V$, the total number of edges (sides), $N$, and the total number of polygons, $F$ (**Fig. 1**). Since three edges meet at every vertex, while each edge links two vertices, $3V = 2N$, we find $F = 2 + \frac{N}{3}$, or, in the limit of $F\rightarrow\infty$, $F = \frac{N}{3}$. In other words, for a large number of polygons, the average number of edges is $N=6$, given that each edge belongs to two polygons [2].

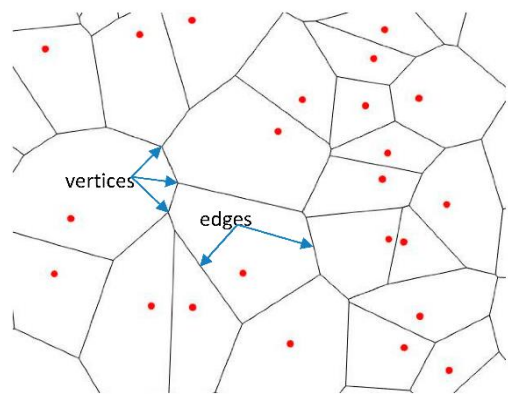

Figure 1. Example of the Voronoi tessellation on a set of points. Red points represent seeds or nuclei [2]. Blue arrows point to the edges of polygons.

For a random (Poisson) set of points at a large 2D plane (so that the *x*- and *y*-coordinates are independent and randomly distributed), the statistical distribution of polygons with different numbers of edges is quite complicated and involves the three-parameter gamma-distributions [12-15]. While analytical solutions have been suggested by Hayen and Quine for *n*=3 [14], and Calca for the general case of any integer *n*≥3 [12-13], these solutions are in the integral form, which requires numerical integration The typical values are presented in **Table 1**.

Table 1. Proportion of polygons with *n*-sides for a random point process, based on [11]

| $n$ | 3 | 4 | 5 | 6 | 7 | 8 | 9 | 10 | 11 | 12 |
|---|---|---|---|---|---|---|---|---|---|---|
| $p_n$ | 0.0113 | 0.1068 | 0.2595 | 0.2947 | 0.1988 | 0.0901 | 0.0296 | 0.0075 | 0.0015 | 0.0002 |

Previously, it has been reported that the value of VE for a random set of points is close to $S \approx 1.71$ [16]; however, more accurate calculations using the data from **Table 1** suggest $S = 1.690 \pm 0.001$. While it is possible to obtain higher values of VE, $S > 1.69$, for certain cases [17], it is not known what is the maximum value, and even whether such a maximum value exists. It is difficult to create a tessellation with a large number of polygons having many edges.

The Voronoi entropy was calculated with the moduli of the software developed at the Department of Physics and Astronomy at the University of California, Irvine (https://www.physics.uci.edu/~foams/do_all.html).

*Parameters and exponents of long-range correlations*

We followed the approach from Refs. [7, 9, 18-19], which present results for hyperuniform systems with partially constrained collective variables [7] and systems of particles interacting with stealthy pair potentials [18-19]. In such systems, the parameter γ characterizing the long-wavelength limit behavior of the autocorrelation function (ACF), $C(\mathbf{R})$, can be introduced as the ratio of the number of constrained degrees of freedom to the total number of degrees of freedom [7]. Note that the hyperuniformity exponent is usually defined as the scaling exponent of the number variance of the system as a function of window size, based on which all hyperuniform systems can be classified into three categories [18-19]. The stealthy systems analyzed here, possess the value of exponent equal to two irrespective of the value of the parameter γ.

The ACF for a random process e.g., a rough solid surface profile $z(x)$, where $z$ is the profile height and $x$ is the horizontal coordinate, is defined by correlating two points $z(x)$ and $z(x + \tau)$ separated by the distance $\tau$

$$C(\tau) = \frac{1}{LR_q^2} \int_0^L [z(x + \tau) - \bar{m}][z(x) - \bar{m}] \mathrm{d}x \qquad (3)$$

where L is the length of the profile, $\bar{m} = \frac{1}{L}\int_0^L z(x)\mathrm{d}x$ is the mean value of $z(x)$, and $R_q^2$ is the standard deviation needed for normalization purposes [20]. Eq. 3 can easily be generalized for the 2D or 3D case of $C(\mathbf{R})$.

The decay of the ACF in the long-range (large wavelength) limit may be exponential, $C(\tau) \sim e^{-\tau/\beta}$, where $\beta$ is the correlation length or the decay may follow the power law,

$$C(\tau) \sim \tau^{\gamma-1}. \tag{4}$$

For a perfect pattern (crystalline material), $\gamma = 1$, and the correlation remains finite in the long-range limit. On the other hand, for a completely random process, $\gamma = 0$.

**Results**

VE was calculated for 14 hyperuniform sets of points obtained from Uche et al. (2004) [7] and from Zhang et al. (2015) [18-19]. The correlation of the VE and the dimensionless parameter $\gamma$ is presented in **Fig. 2**.

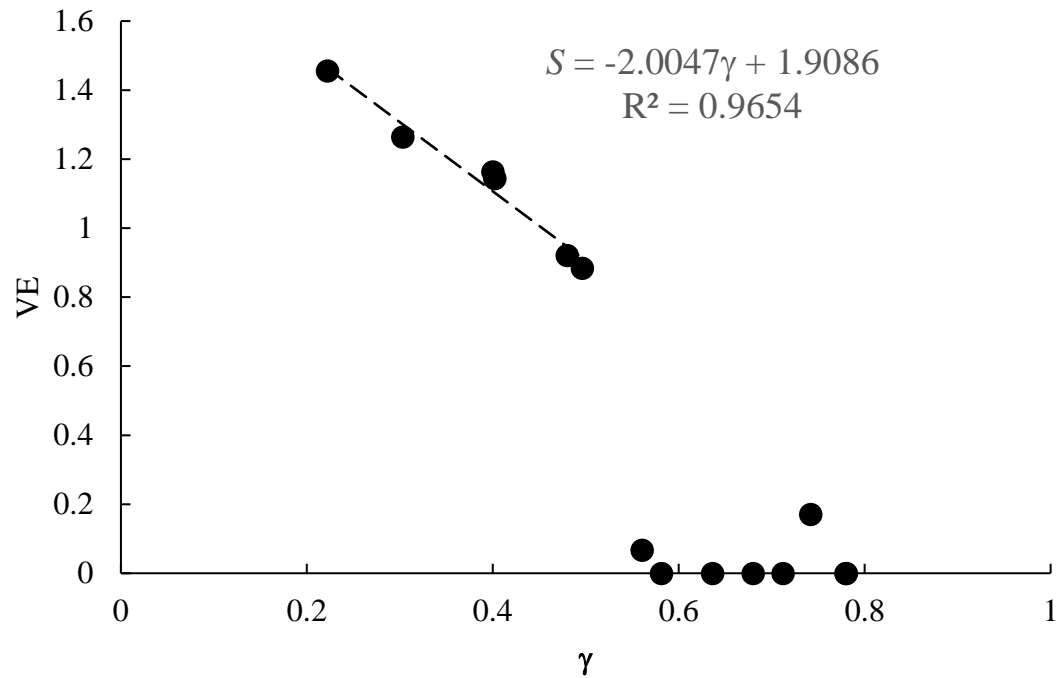

**Figure 2.** Correlation between the parameter $\gamma$ and VE.

It is observed that there is a strong negative correlation between the VE and the exponent parameter up to the value of $\gamma \approx 0.5$. With further increase, a saturation happens and VE abruptly drops to values close to $S=0$. The coefficient of determination for the six data points before the saturation is $R^2 = 0.9654$.

To explain this behavior, one can consider three examples corresponding to $\gamma = 0.22$, $\gamma = 0.58$, $S = 0$, and $\gamma = 0.78$ [7]. **Figure 3** shows the arrangement of points and the Voronoi tessellation for these three cases. The computer program marks polygons with different numbers of vertices by different colors: rectangles (green), pentagons (yellow), hexagons (gray), heptagons (blue), octagons (brown), enneagons (cyan), and decagons (red).

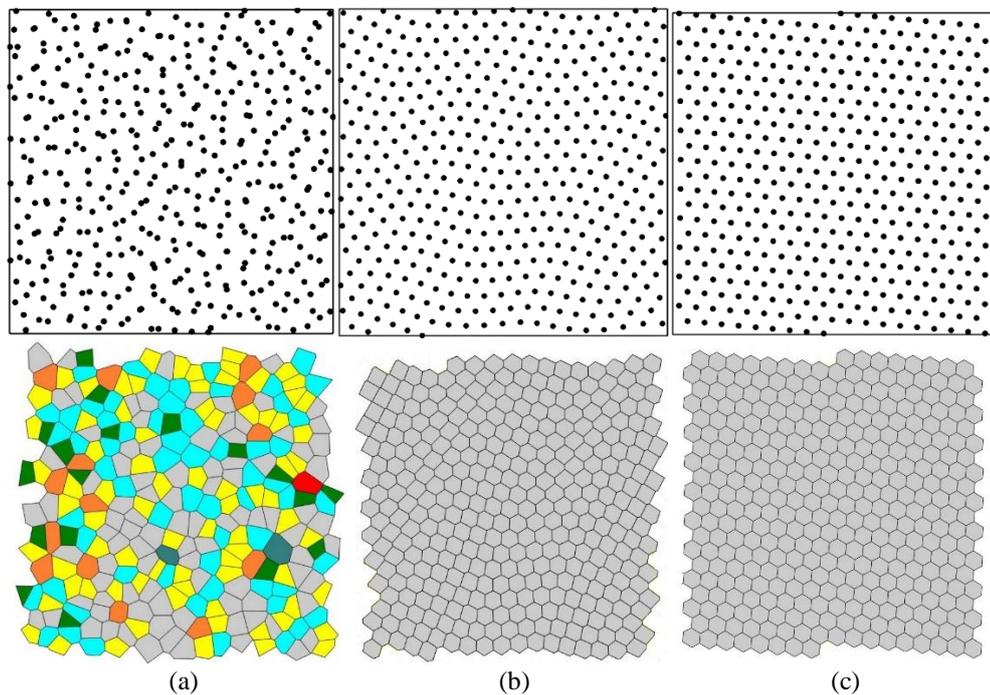

(a)          (b)          (c)
**Figure 3.** (a) Random $\gamma = 0.22$, $S = 1.46$, (b) Wavy Crystalline $\gamma = 0.58$, $S = 0$, and (c) Crystalline $\gamma = 0.78$, $S = 0$.

The first case corresponds to a random point distribution, so the value of $\gamma = 0.22$ is low, while the value of $S = 1.46$ is high. The last case corresponds to a periodic crystalline pattern, so the value of $\gamma = 0.78$ is high, while the value of $S = 0$ is at minimum. The second case corresponds to the "wavy crystalline" pattern, so that the value of $\gamma = 0.58$ is lower than that for the perfect crystallin, $\gamma = 0.58$, however, the Voronoi tessellation results in all polygons being hexagons (although of slightly different sizes). Hence the VE is again at minimum, $S = 0$, which explains the observed saturation.

The abrupt change of VE at $\gamma \cong 0.5$ resembles the behavior of the true thermodynamic entropy under the so-called entropy-driven phase transitions, when the phase-transition is solely driven by the increase in entropy, as it takes place in colloidal crystals [21]. Generally, the VE entropy is very different from the Boltzmann entropy [22]; however, in our research it demonstrated the properties of thermodynamic entropy.

For comparison, the values of the CSM were also calculated using a Matlab program, as $\psi = 0.0216$ for the random (**Fig. 3a**), $\psi = 0.0107$ for the wavy crystalline (**Fig. 3b**), $\psi = 0.0030$ for the crystalline case (**Fig. 3c**); thus, confirming that the crystalline case is more symmetric. Note that the wavy crystalline pattern, depicted in **Figure 3b**, follows the predictions of the Euler equation being constructed from hexagons only. The correlation between the CSM and γ is presented in **Figure 4**. The value of the coefficient of determination is $R^2 = 0.3313$.

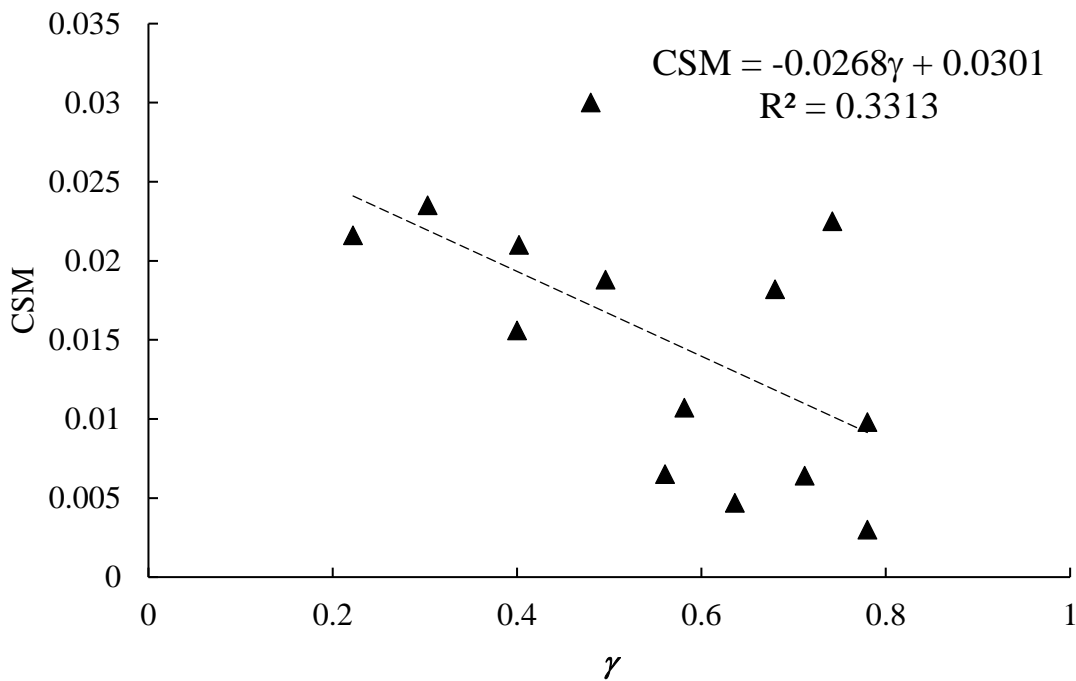

**Figure 4.** Correlation between the hyperuniformity exponent $\gamma$ and CSM.

**Discussion**

Voronoi tessellation is based on the short-range orderliness of the system of points since the number of edges in a polygon depends only on the neighboring polygons. One could expect that it does not capture a long-range order.

The VE is a kind of Shannon entropy applied to the distribution of polygons emerging from the Voronoi tessellation. However, VE possesses specific properties, arising from the geometry of the entire Voronoi tessellation. Any Voronoi tessellation of any planar figure results in a certain distribution of the polygons' probabilities, which is dictated by the topological constraints of forming the planar figure by the polygons. This situation is different from many typical cases of application of the Shannon entropy. For comparison, the Shannon entropy of a text consisting of *m* alphabetic letters (*m*=24 for English) is given by a formula similar to Eq. (1)

$$H = -\sum_{n=1}^{m} p_n \ln(p_n) \qquad (5)$$

where $p_n$ is the relative frequency of the *n*-th letter in the text. In the ideal case of a random text with the equal frequency distribution of letters, $p_n = 1/m$, the value of Shannon entropy is at maximum, $H_{max} = \ln(m)$.

However, the situation is different with the VE, since the polygons are not completely independent from each other, since the tessellation has to keep the general topological properties of the entire tessellation untouched (in particular those, prescribed by the Euler equation). Being a tessellation in one plane constrains the fractions of polygons in a non-obvious way. Therefore, besides pure local short-range interactions, long-range constraints are reflected in the resulting value of the VE. This is the reason why VE correlates with the long-range parameter, such as the critical exponent *γ*.

On the other hand, the VE does not depend on the size or area of polygons. Instead, it takes into account only the number of edges. This explains the saturation (or the abrupt drop) of the correlation dependency in **Figure 2** corresponding to the "wavy crystalline" arrangement of points.

Note also that the wavy crystalline and perfectly crystalline phases cannot be distinguished with VE. One might generalize VE to distinguish these two phases, for example, including a joint distribution of the number of edges and the length of perimeters, or some measure of the anisotropy of Voronoi cells. Another way of generalization of the VE approach may be to consider a more general quantity, such as the Rényi entropy, a generalization of Shannon entropy, which is widely used in ecology as Hill's diversity index. Potentially, VE can be applied to the study of other types of correlated disordered point configurations, such as hard-particle fields and systems with Random sequential adsorption [23].

## Conclusions

We calculated the Voronoi/Shannon entropy (abbreviated VE and denoted by $S$) and the continuous measure of symmetry (abbreviated CSM) for the hyperuniform sets of points with the imposed long-range geometrical correlation and compared the VE and CMS to the values of exponents $\gamma$ of collective density variables characterizing long-range correlations in the system. We found that the VE correlates with the exponents $\gamma$ up to a certain saturation level ($\gamma \cong 0.6$), after which the value of the VE abruptly falls to $S = 0$. We explain this phenomenon by the fact that the distribution of polygons by the number of edges in a given pattern is constrained by being a tessellation within a plane (namely, a surface possessing given topological properties). Therefore, besides pure local short-range correlations, long-range topological constraints are reflected in the resulting value of the VE. This becomes clear if we consider that just the seminal Euler topological Formula predicts that the most abundant polygon in the random Voronoi tessellation is a hexagon. The saturation is explained by the lack of sensitivity of the VE to the size/area of the polygons. This hypothesis will be checked in our future investigations, devoted to the study of the patterns built on a diversity of surfaces.

**Acknowledgement.** This work received no external funding. M. N. recognizes the sabbatical support from the Gale Foundation (Ariel University).

## References

1. Janai, E.; Schofield, A.B.; Sloutskin, E. Non-crystalline colloidal clusters in two dimensions: Size distributions and shapes. Soft Matter 2012, 8, 2924–2929.

2. Bormashenko, Ed.; Frenkel, M.; Vilk, A.; Legchenkova, I.; Fedorets, A. A.; Aktaev, N.; Dombrovsky L. A., Nosonovsky, M. Characterization of Self-Assembled 2D Patterns with Voronoi Entropy. *Entropy* **2018**, *20*(12), 956.

3. Fedorets, A.A.; Bormashenko, E.; Dombrovsky, L.A.; Nosonovsky, M. Droplet clusters: Nature-inspired biological reactors and aerosols. *Philos. Trans. R. Soc. A* **2019**, *377*, 20190121

4. S Shityakov, AS Aglikov, EV Skorb, M Nosonovsky, Voronoi Entropy as a Ligand Molecular Descriptor of Protein–Ligand Interactions, *ACS Omega* **2023**, *8* (48), 46190–46196

5. Botnar, A.; Novokov, O.; Korepanov, O.; et al. "Crystallization Control of Anionic Thiacalixarenes on Silicon Surface Coated with Cationic Poly(ethyleneimine)" *ACS Appl. Mater. Interfaces* 2024 (submitted).

6. Zabrodsky, H.; Peleg, S.; Avnir, D. Symmetry as a continuous feature. *IEEE Trans. Pattern Anal. Mach. Intell.* **1995**, *17*, 1154–1166, DOI: 10.1109/34.476508

7. Uche, O. U.; Stillinger, F. H., and Torquato, S. Constraints on collective density variables: Two dimensions. *Phys. Rev. E* **2004**, *70*, 046122.

8. Frenkel, M.; Fedorets, A. A.; Dombrovsky, L. A.; Nosonovsky, M.; Legchenkova, I.; Bormashenko, E. Continuous Symmetry Measure vs Voronoi Entropy of Droplet Clusters, *J. Phys. Chem. C* **2021**, *125*, 4, 2431–2436

9. Torquato, S. Hyperuniform states of matter. *Phys. Rep.* **2018**, *745*, 1–95

10. Kumar, S., Kurtz, S.K.: Properties of a two-dimensional Poisson-Voronoi tessellation: a Monte-Carlo study. *Mater. Charact.* **1993**, *31(1)*, 55–68.

11. Zhu, H.X.; Thorpe, S.M.; Windle, A.H. The geometrical properties of irregular two-dimensional. *Phil. Mag. A* **2001**, *81*, 2765–2783.

12. Calka, P.: The explicit expression of the distribution of the number of sides of the typi-cal Poisson Voronoi cell. *Advances in Applied Probability* **2003**, *35*(4), 863-870.

13. Calka, P. Precise Formulae for the Distributions of the Principal Geometric Characteristics of the Typical Cells of a Two-Dimensional Poisson-Voronoi Tessellation and a Poisson Line Process, *Adv. Appl. Probability* **2003**, *35*(3), pp. 551-562.

14. Hayen, A.; Quine, M. The proportion of triangles in a Poisson-Voronoi tessellation of the plane. *Adv. Appl. Prob. (SGSA)* **2002**, *32*, 67–74.

15. Jarai-Szabo, F.; Zoltan, N. On the size distribution of Poisson Voronoi cells. *Phys. A* **2007**, *385*, 518–526.

16. Limaye, A.V.; Narhe, R.D.; Dhote, A.M.; Ogale, S.B. Evidence for convective effects in breath figure formation on volatile fluid surfaces. *Phys. Rev. Lett.* **1996**, *76*, 3762–3765.

17. Frenkel, M.; Legchenkova, I.; Shvalb, N.; Shoval, S.; Bormashenko, E. Voronoi Diagrams Generated by the Archimedes Spiral: Fibonacci Numbers, Chirality and Aesthetic Appeal. *Symmetry* **2023**, *15*, 746. https://doi.org/10.3390/sym15030746

18. Zhang, G.; Stillinger, F. H.; Torquato, S. Ground states of stealthy hyperuniform potentials: I. Entropically favored configurations, *Phys Rev E* **2015**, *92*, 022119.

19. Zhang, G.; Stillinger, F. H.; Torquato, S. Ground states of stealthy hyperuniform potentials: II. Stacked-slider phases, *Phys Rev E* **2015**, *92*, 022120.

20. Aglikov, A. S., et al., Topological Data Analysis of Nanoscale Roughness of Layer-by-Layer Polyelectrolyte Samples Using Machine Learning, *ACS Applied Electronic Materials* **2023**, *5* (12), 6955–6963.

21. Dijkstra, M. Entropy-Driven Phase Transitions in Colloids: From spheres to anisotropic particles, *Advances in Chemical Physics*, Vol. 156 (Eds. Rice, S.A.; Dinner A. R) Wiley, 2014)

22. Ben-Naim A. Entropy, Shannon's Measure of Information and Boltzmann's H-Theorem, *Entropy* 2017, 19(2), 48

23. Kubala, P., Cieśla, M., and Ziff, R. M. Random sequential adsorption of particles with tetrahedral symmetry, Phys. Rev. E 2019, 100, 052903